\documentstyle[prl,aps]{revtex}
\newcommand{\be}{\begin{equation}}
\newcommand{\ee}{\end{equation}}
\newcommand{\ba}{\begin{eqnarray}}
\newcommand{\ea}{\end{eqnarray}}

\def\f{\phi}

\def\m{\mu}

\begin{document}
\draft
\title{Comment on ``On spin-1 massive particles coupled to a 
Chern-Simons field''}
\author{O.M. Del Cima$^a$\thanks{\tt oswaldo@cbpf.br}, 
D.H.T. Franco$^{b}$\thanks{\tt dfranco@cbpf.br}, 
J.A. Helay\"el-Neto$^{a,b}$\thanks{\tt helayel@cbpf.br} and 
O. Piguet$^c$\thanks{{\tt piguet@npd.ufes.br} and 
{\tt piguet@cce.ufes.br}}     }
\address{$^a${\it Centro Brasileiro de Pesquisas F\'\i sicas (CBPF),}\\
Departamento de Teoria de Campos e Part\'\i culas (DCP), \\
Rua Dr. Xavier Sigaud 150 - 22290-180 - Rio de Janeiro - RJ - Brazil.\\
$^b${\it Universidade Cat\'olica de Petr\'opolis (UCP),} \\ 
Grupo de F\'\i sica Te\'orica, \\ 
Rua Bar\~ao do Amazonas 124 - 25685-070 - Petr\'opolis - RJ - Brazil.\\
$^c${\it Universidade Federal do Esp\'\i rito Santo (UFES),}\\
CCE, Departamento de F\'\i sica, \\
Campus Universit\'ario de Goiabeiras - 29060-900 - Vit\'oria - ES - Brazil.   }
\date{\today}
\maketitle
\begin{abstract}

In this comment we discuss some serious inconsistencies 
presented by Gomes, Malacarne and da Silva in their paper, 
Phys.Rev. D60 (1999) 125016 (hep-th/9908181). 
\end{abstract}
\pacs{PACS numbers: 11.10.Gh 11.10.Kk 11.15.-q 11.80.-m 
\hspace{4,05cm}CBPF-NF-065/99}


In the paper of Ref.\cite{Gomes}, Gomes, Malacarne and 
da Silva, set up some conclusions about the dynamics 
and interactions between charged vector bosons ($\f_\m$) and 
the gauge field ($A_\m$) in 3 dimensions. Besides, they also 
discuss the issue of 
1-loop renormalizability. This comment is devoted to 
point out some inconsistencies in their misleading analysis; 
mainly, we criticize the way the authors heavily use the 
Ward identities to ``ensure'' 1-loop renormalizability 
for a model which is not even unitary at tree-level
thanks to the violation of the Froissart-Martin bound. 

According to the results previously worked out in the papers 
of Refs.\cite{DelCima1,DelCima2,Tese}, following a line 
initiated by \cite{Hagen}, it is known that charged vector 
fields minimally (or non-minimally) coupled to a gauge field 
display severe problems in what concerns the quantum-mechanical 
consistency of the model. To be more specific: unitarity is 
jeopardized by complex massive vector fields, regardless 
the mass is introduced via a Proca or a Chern-Simons term, 
as we shall clarify below. 

The authors of Ref.\cite{Gomes} claim that, even if a Proca 
term assigns mass to the charged vector field, the 1-loop 
renormalizability of the model is guaranteed by virtue of the 
identity of Eq.(16) in their paper. However, the use of such 
an identity in the calculation of 1-loop graphs such as 
vacuum-polarization and the 4-point function for the 
Chern-Simons field is not appropriate to reduce the superficial 
degree of divergence, for there is not reason to set the 
momenta associated to the matter-field lines in the 3-vertex 
equal to zero. Our remark is that there is no way to tame the 
ultraviolet divergences brought about by the Proca term. On 
the other hand, following the results of 
\cite{DelCima1,DelCima2,Tese}, the dynamical induction 
of a Proca term always takes place for topologically massive 
complex 
vector fields. The criterium that is neglected in 
the analysis of the work of Ref.\cite{Gomes} (the same 
criticism applies to the work by Bezerra de Mello and 
Mostepanenko \cite{Bezerra}) is the lack of 
reference to the 
Froissart-Martin bound in 3 dimensions \cite{Chaichian,Chadan}, 
which is of the type ``$s~{\rm ln}s$'' for the total scattering 
squared amplitude in a Compton-like process. 
Though it is not very 
evident, the actually serious problem of the massive Proca 
complex vector field is that it leads to a clear violation of 
the Froissart-Martin unitarity bound in 3 dimensions, yielding 
an upper bound of the form ``$s^2$'' \cite{DelCima1}. Had we 
started with a topologically massive complex vector field, unitarity 
would apparently be respected through the Froissart-Martin 
bound \cite{Chaichian,Chadan}, since an upper bound of 
the type ``$s^0$'' \cite{DelCima1} shows up for that case; 
nevertheless, a Proca term 
($\f^*_\m\f^\m$) is 
always radiatively induced and a non-unitary bound ``$s^2$'' 
drops out \cite{DelCima1,DelCima2,Tese}.

Our comment sets out to raise the question whether it is 
sensible to consider the massive charged vector model beyond 
the tree-approximation, once, as stated above, the unitarity 
bound is clearly violated at that approximation. Usually, we 
draw our attention to the renormalizability and unitarity by 
taking into account power-counting, counter-terms, Ward 
identities and the positivity of the states in the Hilbert 
space. Though these are necessary requirements to be fulfilled, 
some additional criteria ought to be checked, such as the 
validity of the Froissart-Martin bound. The class of models 
discussed in Refs.\cite{Gomes,Bezerra} is a good warning 
example for what we have just mentioned: though the analysis 
of propagators and power-counting seems to point out to a 
healthy model in the case of Maxwell-Chern-Simons theory for 
the complex vector field, the induction of the Proca mass 
breaks the unitarity bound and we believe it is not sensible 
to go beyond tree-level, or, in short, to second-quantize 
such a model.

We should also stress that the introduction of a 
gauge-invariant non-minimal magnetic coupling, which in the 
Proca case is non-renormalizable, does not restore the 
Froissart-Martin bound in that case of 
Maxwell-Chern-Simons-Proca model for the charged vector field, 
as it was attained in Ref.\cite{DelCima1}.

To end our short comment, we conclude that, besides the lack 
of power-counting renormalizability, unitarity is the 
key ingredient to rule out the theory of massive charged 
vector fields coupled to a gauge field in 3 dimensions as a 
fundamental field theory, therefore, the results of Refs.\cite{DelCima1,DelCima2} turn those of Ref.\cite{Gomes} 
useless.


\begin{references}
\bibitem{Gomes} M. Gomes, L.C. Malacarne and A.J. da Silva, 
Phys.Rev. D60 (1999) 125016, hep-th/9908181. 
\bibitem{DelCima1} O.M. Del Cima, Mod.Phys.Lett. A9 (1994) 
1695.
\bibitem{DelCima2} O.M. Del Cima and F.A.B. Rabelo de Carvalho, 
Int.J.Mod.Phys. A10 (1995) 1641.
\bibitem{Tese} O.M. Del Cima, M.Sc. Thesis: 
{\it Quantum aspects of complex vector fields in $D$$=$$3$}, in 
portuguese, CBPF-DCP (July 1993) - Rio de Janeiro - Brazil.
\bibitem{Hagen} C.R. Hagen, P. Panigrahi and S. Ramaswamy, 
Phys.Rev.Lett. 61 (1988) 389; C.R. Hagen and S. Ramaswamy, 
Phys.Rev. D41 (1990) 1920 and Phys.Rev. D42 (1990) 3524; 
C.R. Hagen, Int.J.Mod.Phys. A6 (1991) 3119.
\bibitem{Bezerra} E.R. Bezerra de Mello and V.M. Mostepanenko, 
Int.J.Mod.Phys. A14 (1999) 271.
\bibitem{Chaichian} M. Chaichian and J. Fischer, 
Nucl.Phys. B303 (1988) 557; M. Chaichian, J. Fischer 
and Yu.S. Vernov, Nucl.Phys. B383 (1992) 151.
\bibitem{Chadan} K. Chadan, N.N. Khuri, A. Martin and 
T.T. Wu, Phys.Rev. D58 (1998) 025014, hep-th/9805036.
\end{references}
\end{document}